\pdfoutput=1
\documentclass[envcountsame,orivec,fleqn]{llncs}

\usepackage{cite}
\usepackage{hyperref}
\usepackage[capitalise]{cleveref}
\usepackage{amsmath}
\usepackage[english]{babel}
\usepackage{latexsym}
\usepackage{color}
\usepackage[usenames,dvipsnames]{xcolor}
\usepackage{amssymb}	
\usepackage{verbatim}   
\usepackage{ifthen}
\usepackage{xfrac}      
\usepackage{graphicx}   
\usepackage{wrapfig}
\usepackage[plain]{algorithm}
\usepackage{algpseudocode}
\usepackage{paralist}
\usepackage{array}
\usepackage{multirow}
\usepackage{rotating}
\usepackage{url}

\makeatletter
\newlength{\@auxiliaryquotewidth}

\newenvironment{labelledquote}
               {\setlength{\@auxiliaryquotewidth}{\textwidth}\addtolength{\@auxiliaryquotewidth}{-\leftmargin}\list{}{\rightmargin1.5\leftmargin}%
                \item\relax\refstepcounter{equation}\makebox[-4pt][l]{\makebox[\@auxiliaryquotewidth][r]{(\theequation)}}}
               {\endlist}

\makeatother

\Crefname{equation}{}{}

\definecolor{lightblue}{RGB}{210,210,225}
\definecolor{lightred}{RGB}{225,210,210}
\definecolor{lightgreen}{RGB}{210,225,210}
\definecolor{lightyellow}{RGB}{225,222,200}
\definecolor{lightpurple}{RGB}{225,210,225}
\definecolor{darkerred}{RGB}{64,0,0}
\definecolor{darkred}{RGB}{128,0,0}
\definecolor{darkblue}{RGB}{0,0,128}
\definecolor{darkgreen}{RGB}{0,128,0}
\definecolor{darkpurple}{RGB}{128,0,128}

\newcommand{\colorpar}[3]{\colorbox{#1}{\parbox{#2}{#3}}}

\newcommand{\marginremark}[4]{\marginpar{\colorpar{#2}{\linewidth}{\color{#1}\tiny{[#3]~ #4}}}}

\newcommand{\citeref}[2]{\href{#1}{#2}~\cite{#1}}

\makeatletter
\def\THICKhrulefill{\leavevmode \leaders \hrule height 5pt\hfill \kern \z@}
\makeatother

\iftrue

  \newcommand{\remarkHH}[1]{\marginremark{darkblue}{lightblue}{HH}{#1}}
  \newcommand{\remarkPRD}[1]{\marginremark{darkred}{lightred}{PRD}{#1}}
\else
  \newcommand{\remarkPRD}[1]{}
  \newcommand{\remarkCEB}[1]{}
  \newcommand{\remarkHH}[1]{}
\fi

\titlerunning{Facets of Software Doping}
\authorrunning{G.~Barthe \and P.R.~D'Argenio \and B.~Finkbeiner \and H.~Hermanns}

\begin{document}
\title{
Facets of Software Doping%
	\thanks{This work is partly supported by the ERC Grants 683300 (OSARES) and 695614 (POWVER), by the Sino-German CDZ project 1023 (CAP), by ANPCyT PICT-2012-1823, by SeCyT-UNC 05/BP12 and 05/B497, and by the Madrid Region project S2013/ICE-2731 N-GREENS Software-CM.%
	}%
}%
\author{
	Gilles Barthe\inst{1}    \and
	Pedro R.~D'Argenio\inst{2} \and\\
	Bernd Finkbeiner\inst{3} \and
	Holger Hermanns\inst{3}
}%
\institute{IMDEA Software
  \and
  FaMAF, Universidad Nacional de C\'ordoba -- CONICET
  \and 
  Saarland University -- Computer Science, Saarland Informatics Campus
}
\maketitle

\begin{abstract}
This paper provides an informal discussion of the 
formal aspects of software doping.
\end{abstract}

\section{Introduction}

Software is the great innovation enabler of our times. Software runs
on hardware. Usually, software is licensed to the hardware owner,
instead of being owned by her.  And while the owner is in full
physical control of the hardware, she usually has neither physical nor
logical control over the software. That software however does not
always exploit the offered functionality of the hardware in the best
interest of the owner. Instead it may be tweaked in various manners,
driven by interests different from those of the owner or of society.
This situation may be aggravated if the software is not
running on local hardware but remotely (e.g. in the cloud) since the
software user has now little or no control of its execution.

There is a manifold of facets to this phenomenon, summarised as
software doping. It becomes more widespread as software is embedded in
ever more devices of daily use. Yet, we are not aware of any
systematic investigation or formalisation from the software
engineering perspective.

This paper reviews known real cases of software doping, and provides a
conceptual account of characteristic behaviour that distinguishes
doped from clean software.

\section{Software Doping in the Wild}

The simplest and likely most common example of software doping is that
of   \citeref{http://www.wasteink.co.uk/epson-firmware-update-compatible-problem/}{ink printers} refusing to work when supplied with a toner or ink cartridge of a \citeref{https://conversation.which.co.uk/technology/printer-software-update-third-party-printer-ink/}{third party manufacturer}, albeit being technically compatible. More subtle variations of this kind of doping just issue a warning
message about the \citeref{http://uk.pcmag.com/printers/60628/opinion/the-secret-printer-companies-are-keeping-from-you}{risk of using a ``foreign'' cartridge}. In the same
vein, it is known that printers \citeref{http://www.slate.com/articles/technology/technology/2008/08/take_that_stupid_printer.html}{emit ``low toner'' warnings} earlier than needed, so as to drive or force the customer into replacing cartridges prematurely. 
Similarly, cases are known where \citeref{https://nctritech.wordpress.com/2010/01/26/dell-laptops-reject-third-party-batteries-and-ac-adapterschargers-hardware-vendor-lock-in/}{laptops refuse to charge} the battery if connected to a third-party charger. 

Characteristic for these examples is that the functionality in question is in the interest of the device manufacturer, but against the customer interest. However, there are also 
variations of software doping  that can be considered to be in the
interest of the customer, but not in the interest of society: In the
automotive sector,
\citeref{https://en.wikipedia.org/wiki/Chip_tuning}{``chip-tuning''} is a
remarkable variation of the software doping phenomenon, where the owner initiates a
reprogramming of some of the vehicle's electronic control units (ECU)
so as to change the vehicle characteristics with respect to power,
emissions, or fuel consumption. By its nature, chip-tuning appears to
be in the owner's interest, but it may well be against the interest of
society, for instance if legally-defined and thus built-in speed
limitations are overridden. Examples include \citeref{https://rideapart.com/articles/hows-work-ecu-tuning}{scooters} and
\citeref{http://www.ebiketuning.com/comparison/bionx-tuning.html}{electric bikes}\cite{http://electricbikereview.com/community/threads/how-to-get-access-to-your-bionx-console-code-menu-codelist-included.519/}.

Some cases of software doping are clearly neither in the interest of
the customer, nor in the interest of society. This includes as
prominent examples the \citeref{https://en.wikipedia.org/wiki/Volkswagen_emissions_scandal}{exhaust emission scandal of Volkswagen} (and other manufacturers). Here, the exhaust software was manufactured  in
such a way that it heavily polluted the environment, unless the
software detected the car to be fixed on the particular test setup
used to determine the NOx footprint data officially published.

The same sort of behaviour has been reported in the context of \citeref{http://forums.appleinsider.com/discussion/158782/galaxy-s-4-on-steroids-samsung-caught-doping-in-benchmarks}{smart
phone designs}, where software was tailored to perform better when
detecting it was running a certain benchmark, and otherwise running in
lower clock speed. Another smart phone case, \citeref{https://www.theguardian.com/money/2016/feb/05/error-53-apple-iphone-software-update-handset-worthless-third-party-repair}{disabling the phone} via a
software update after ``non-authorized'' repair, has later been \citeref{https://techcrunch.com/2016/02/18/apple-apologizes-and-updates-ios-to-restore-iphones-disabled-by-error-53/}{undone}. Often, software doping is a part of a lock-in strategy: The customer gets \emph{locked-in} on the manufacturer or unit-supplier for products, maintenance and services \cite{ecojournal1989}.

\section{Characterising Software Doping}

It is  difficult to come up with a crisp characterisation of what constitutes software
doping. Nevertheless we consider it a worthwhile undertaking to explore this issue, with the intention to eventually enable a formal characterisation of software doping. That characterisation can be the nucleus for formulating and enforcing rigid requirements on embedded software driven by public interest, so as to effectively ban software doping.
In order to sharpen our intuition, we offer the
following initial characterisation attempt.

\begin{labelledquote}\label{dope:intuition}
  A software system is doped if the manufacturer has included a hidden
  functionality in such a way that the resulting behaviour
  intentionally favors a designated party, against the interest of
  society or of the software licensee.
\end{labelledquote}

So, a doped software induces behaviour that can not be justified by
the interest of the licensee or of society, but instead serves another
usually hidden interest. It thereby favors a certain brand, vendor,
manufacturer, or other market participant. This happens intentionally, and not by accident.

However, the question whether a certain behaviour is intentional or not  is very  difficult to  decide. To illustrate this, we recall that the above mentioned iPhone-6 case,
where \citeref{https://www.theguardian.com/money/2016/feb/05/error-53-apple-iphone-software-update-handset-worthless-third-party-repair}{``non-authorized'' repair rendered the phone unusable} after an iOS update, seemed to be intentional when it surfaced, but was actually tracked down to a software glitch of the update and fixed later. Notably, if the iOS designers would have had the particular intention to mistreat licensees who went elsewhere for repair, the same behaviour could well have qualified as software doping in the above sense~\cref{dope:intuition}.

As a result, we will look at software doping according to the above characterisation, but without any attempt to take  into account  considerations of intentionality.

In the sequel, we shall investigate this phenomenon by synthetic
examples, that however are directly inspired by the real cases of
software doping reviewed above.


\subsection{Doping by discrimination}

\newcommand{\var}[1]{\ensuremath{\textit{#1}}}

\begin{wrapfigure}[24]{r}{6.9cm}
  \small\vspace{-2.5em}%
  \begin{algorithmic}
  \Procedure{Printer}{\var{cartridge\_info}}
  \State \Call{read}{\var{document}}
  \While {$\Call{pagesToPrint}{\var{document}}>0$}  
    \State \Call{read}{\var{paper\_available?}}
    \If {$\neg\var{paper\_available?}$}
      \State \Call{turnOn}{\var{alert\_signal}}
      \State \Call{waitUntil}{\var{paper\_available?}}
      \State \Call{turnOff}{\var{alert\_signal}}
    \EndIf
    \State \Call{printNextPage}{\var{page\_out},\var{document}}
  \EndWhile
  \EndProcedure
  \end{algorithmic}
  \caption{A simple printer.}\label{fig:printer:general}
  \vspace{1em}
  \begin{algorithmic}
    \Procedure{Printer}{\var{cartridge\_info}}
    \If {$\Call{brand}{\var{cartridge\_info}} = \var{my-brand}$}
    \State ($\cdots$ same code as \cref{fig:printer:general} $\cdots$)
    \Else
    \State \Call{turnOn}{\var{alert\_signal}}
    \EndIf
  \EndProcedure
  \end{algorithmic}
  \caption{A doped printer.}\label{fig:printer:doped}
\end{wrapfigure}%
Think of a program as a function that accepts some initial parameters
and, given (partial) inputs, it produces (partial) outputs. As an
example, (an abstraction of) the embedded software in a printer is
given in \cref{fig:printer:general}.
The program \textsc{Printer} has the parameter
\var{cartridge\_info} (which is not yet used within the
function), two input variables (\var{document} and
\var{paper\_available?}) and two output variables (\var{alert\_signal}
and \var{page\_out}).

%
A printer manufacturer may manipulate this program in order to favor
its own cartridge brand. An obvious way is displayed in 
\cref{fig:printer:doped}. This is a sort of discrimination based on parameter values.  
Therefore, a first formal approach to characterising a program as 
\emph{clean} (or \emph{doping-free}) is that it should behave in a similar way for all
parameter values, where by \emph{similar behaviour} we mean that the
visible output should be the same for any given input in two different
instances of the same (parameterized) program.  Obviously, ``all
parameter values'' refers to all values within a given domain.  In the
case of the printer, we expect that it works with any
\emph{compatible} cartridge.  Such compatibility domain defines a
first scope within which a software is evaluated to be clean or
doped. So, we could say the following.

\begin{labelledquote}\label{dope:def:i}
  A program is \emph{clean} (or \emph{doping-free}) if for every standard
  parameter it exhibits the same visible outputs when supplied with
  the same inputs.
\end{labelledquote}
\noindent%
Under this view, the program of \cref{fig:printer:doped} is indeed
doped.  Also, note that this characterisation entails the existence of
a contract which defines the set of standard parameters.

\subsection{Doping vs. extended functionality}

We could imagine, nonetheless, that the printer manufacturer may like
to provide extra functionalities for its own product which is outside
of the standard for compatibility.  For instance (and for the sake of
this discussion) suppose the printer manufacturer develops a new file
format that is more efficient at the time of printing, but this
requires some new technology on the cartridge.  The manufacturer still
wants to provide the usual functionality for standard file formats
that works with standard compatible cartridges and comes up with the
program of \cref{fig:printer:nondoped}.
\begin{figure}[t]
  \small
  \begin{algorithmic}
    \Procedure{Printer}{\var{cartridge\_info}}
    \State \Call{read}{\var{document}}
    \If {${\neg \Call{newType}{\var{document}}}{} \lor {}{\Call{supportsNewType}{\var{cartridge\_info}}}$}
    \State ($\cdots$ proceed to print as in \cref{fig:printer:general} $\cdots$)
    \Else
    \State \Call{turnOn}{\var{alert\_signal}}
    \EndIf
    \EndProcedure
  \end{algorithmic}
  \caption{A clean printer.}\label{fig:printer:nondoped}
\end{figure}
Notice that this program does not conform to the specification of a
clean program given by \cref{dope:def:i} since it behaves
differently when a document of the new (non-standard) type is given.
This is clearly not in the spirit of the program in
\cref{fig:printer:nondoped} which is actually conforming to the
standard specification.  Thus, we relax the previous characterisation
and only require that two instances of the program behave similarly if
the provided inputs adhere to some expected standard.  Therefore we
propose the following weaker notion of clean program:

\begin{labelledquote}\label{dope:def:ii}
  A program is \emph{clean} if for every standard parameter it
  exhibits the same visible outputs when supplied with any possible
  input complying with a given standard.
\end{labelledquote}

This characterisation is based on a comparison of the behaviour of two
instances of a program, each of them responding to different parameter
values.  A second, different characterisation may instead require to compare a
reference specification capturing the essence of clean behaviour against
any possible instance of the program.  The first approach seems more
general than the second one in the sense that the specification could
be considered as one of the possible instances of the (parameterized)
program.
However, the second characterisation is still reasonable and it could
turn to be equivalent to \cref{dope:def:ii} under mild conditions
(namely, under behavioural equivalence.)

\subsection{Doping by switching}

\begin{wrapfigure}[7]{r}{5.6cm}
  \small%
  \vspace{-2.4em}%
  \begin{algorithmic}
    \Procedure{EmissionControl}{{}}
      \State \Call{read}{\var{throttle}}
      \State \var{def\_dose} = \Call{SCRModel}{\var{throttle}}
    \EndProcedure
  \end{algorithmic}
  \caption{A simple emission control.}\label{fig:ecu:general}
\end{wrapfigure}
%
Let us draw the reader's attention to a different facet of software doping. We consider  the ECU of a diesel
vehicle, in particular its exhaust emission control module. For diesel engines, the controller
injects a certain amount of a specific fluid (an aqueous urea solution) into the exhaust pipeline in order to lower NOx emissions.
We
 simplify this control problem to a minimal toy example. In \cref{fig:ecu:general} we display  a function that reads the
\var{throttle} position and calculates which is the dose of diesel
exhaust fluid (DEF) that should be injected to reduce the NOx emission
(this is stored in \var{def\_dose}).  Variable \var{throttle} is an
input variable while, though \var{def\_dose} is an output variable, it
is not the actual visible output.  The actual visible output is the
NOx emission measured at the end of the exhaust system.  Therefore,
the behaviour of this system needs to be analyzed through testing.  In
this setting, we may only consider the standard input behaviour as the
one defined in the laboratory emission tests.

\begin{wrapfigure}[10]{r}{7.6cm}
  \small\vspace{-2.6em}%
  \begin{algorithmic}
    \Procedure{EmissionControl}{{}}
      \State \Call{read}{\var{throttle}}
      \If {$\var{throttle}\in\var{throttleTestValues}$}
      \State \var{def\_dose} = \Call{SCRModel}{\var{throttle}}
      \Else
      \State \var{def\_dose} = \Call{alternateSCRModel}{\var{throttle}}
      \EndIf
    \EndProcedure
  \end{algorithmic}
  \caption{A doped emission control.}\label{fig:ecu:doped}
\end{wrapfigure}
%
The Volkswagen emission scandal arose precisely because their software was instrumented  so that it works as expected
\citeref{https://events.ccc.de/congress/2015/Fahrplan/events/7331.html}{\emph{only if} operating in or very close to the lab testing conditions}. For our simplified example, this behaviour is exemplified by  the algorithm of \cref{fig:ecu:doped}. Of course, the real case was less simplistic. 
%
%
Notably, a software like this one still meets the caracterization of
\emph{clean} given in \cref{dope:def:ii}.  However, it is
intentionally programmed to defy the regulations when being unobserved and
hence it falls directly within our intuition of what a doped software
is (see \cref{dope:intuition}).

The spirit of the emission tests is to verify that the amount of NOx
in the car exhaust gas does not exceed a given threshold \emph{in
  general}.  Thus, one would expect that if the input values of the
\textsc{EmissionControl} function deviates within ``reasonable
distance'' from the \emph{standard} input values provided during the
lab emission test, the amount of NOx found in the exhaust gas is
still within the regulated threshold, or at least it does not exceed it
more than a ``reasonable amount''.
Similar rationale could be applied for regulation of other systems
such as
speed limit controllers in scooters and electric bikes.
Therefore, we propose this alternative characterisation:
\begin{labelledquote}\label{dope:def:iii}
  A program is \emph{clean} if for every standard parameter,
  whenever it is supplied with any input (being it complying to the
  standard or not) that deviates within ``reasonable distance'' from a
  given standard input, it exhibits a visible output which does not
  deviate beyond a ``reasonable distance'' from the specified output
  corresponding to such standard input.
\end{labelledquote}

The ``reasonable distances'' are values that should be provided
(together with the notion of distance) and are part of the contract
that ensures that the software is clean. 
Also, the limitation to this ``reasonable distance'' has to do with
the fact that, beyond it, particular requirements (e.g. safety) may
arise.  For instance, a smart battery may decide to stop accepting charge if
the current emitted by  a standardized but foreign charger is higher
than ``reasonable'', but it may still proceed in case it is instead 
dealing with a charger of the same brand for which it may know that it can resort to a customized protocol allowing ultra-fast charging in a safe manner.

These 'reasonable distances'  need to come with application-specific metrics on possible input and output values. Since these metrics are often related to real physical quantities, the metric spaces might be continuous. They might also be discrete, or superpositions of both.  

Characterisation \cref{dope:def:iii} also plays a role when inputs and
outputs cannot be precisely defined.  This situation arises in cases
like the exhaust emission system and almost any embedded system: input
and output values will be as precise as sensors and actuators allow.
In this case, the ``reasonable distance'' is going to be defined
according to the precision of these devices.

\section{Concluding remarks}

This paper has reviewed facets of software doping. Starting off from
real examples a first intuitive characterisation of software doping
was derived. We then discussed a sequence of -- still informal ---
definitions of absence of software doping.

We are currently working on the formalisation of these definitions.
We expect that many definitions will fall in the general class of
hyperproperties~\cite{ClarksonS08}---informally, hyperproperties are
sets of sets of program executions and capture behaviours of multiple
runs of a program---which encompasses continuity~\cite{ChaudhuriGL10}
and non-interference~\cite{GoguenMeseguer82:ieeesp,VolpanoIS96:jcs}.
These formal characterisations are expected to help to understand
better the requirements on embedded software imposed by public
interest, hence providing a framework to specify contracts or
regulations pertaining to such technology, and to rigorously
discriminate between doping and reasonably acceptable deviations from
the normal behaviour. They will also help clarify the specificities of
software doping with respect to malware, software sabotage, and
substitution attacks that have been studied in the context of
security~\cite{Schneier:2015}. Furthermore, rigorous definitions will
provide the necessary foundations for developing analysis methods
(verification or testing) against doping.

\begin{sloppypar}
\bibliographystyle{splncs03}
\bibliography{softwaredoping}
\end{sloppypar}

\end{document}